\documentclass[12pt]{iopart}

\usepackage{iopams}
\usepackage{graphicx}
\usepackage{epstopdf}
\usepackage[]{cite}

\begin{document}

\title[Tunable ground states in helical $p$-wave Josephson junctions]{Tunable ground states in helical $p$-wave Josephson junctions}

\author{Qiang Cheng$^1$, Kunhua Zhang$^2$, Dongyang Yu$^3$, Chongju Chen$^4$ Yinhan Zhang$^5$ and Biao Jin$^4$}

\address{$^1$ School of Science, Qingdao Technological University, Qingdao, Shandong 266520, China}
\address{$^2$ ICQD, Hefei National Laboratory for Physical Sciences at Microscale, University of Science and Technology of China, Hefei, Anhui 230026, China}
\address{$^3$ Department of Physics, Renmin University of China, Beijing 100872, China}
\address{$^4$ School of Physics, University of Chinese Academy of Sciences, Beijing 100049, China}
\address{$^5$ International Center for Quantum Materials, Peking University, Beijing 100871, China}

\ead{chengqiang07@mails.ucas.ac.cn}
\ead{zhangyinhan2008@gmail.com}
\ead{biaojin@ucas.ac.cn}
\vspace{10pt}
\begin{indented}
\item[]
\end{indented}

\begin{abstract}
We study new types of Josephson junctions composed of helical $p$-wave superconductors with $k_{x}\hat{x}\pm k_{y}\hat{y}$ and $k_{y}\hat{x}\pm k_{x}\hat{y}$-pairing symmetries using quasiclassical Green's functions with the generalized Riccati parametrization. The junctions can host rich ground states: $\pi$ phase, $0+\pi$ phase, $\varphi_{0}$ phase and $\varphi$ phase. The phase transition can be tuned by rotating the magnetization in the ferromagnetic interface. We present the phase diagrams in the parameter space formed by the orientation of the magnetization or by the magnitude of the interfacial potentials. The selection rules for the lowest order current which are responsible for the formation of the rich phases are summarized from the current-phase relations based on the numerical calculation. We construct a Ginzburg-Landau type of free energy for the junctions with ${\bf{d}}$-vectors and the magnetization, which not only reveals the interaction forms of spin-triplet superconductivity and ferromagnetism but also can directly leads to the selection rules. In addition, the energies of the Andreev bound states and the novel symmetries in the current-phase relations are also investigated. Our results are helpful both in the prediction of the novel Josephson phases and in the design of the quantum circuits.
\end{abstract}

\pacs{74.50.+r, 74.45.+c, 76.50.+g}
%
\noindent{\it Keywords}: Josephson junctions, helical superconductivity, ferromagnetism
%
%
%
%

\section{Introduction}
Josephson junctions have been subjected to continuously growing interests because of rich ground states in these systems and their potential applications in superconducting electronics \cite{Ioffe,Blatter,Blais,Goldobin5,Linder}. The ground states can be classified into $0$ phase, $\pi$ phase, $\varphi_{0}$ phase and $\varphi$ phase according to the number and the position of the energy minimum within $2\pi$ interval of the superconducting phase $\phi$ across the junctions. The junctions in the $0$ phase and $\pi$ phase, which have been realized experimentally \cite{Jiang, Ryazanov, Robinson, Shelukhin}, have an energy minimum at $\phi=0$ and $\phi=\pi$ \cite{Golubov, Bulaevskii}, respectively, while the $\varphi_{0}$ junctions have a single energy minimum at $\phi=\varphi_{0}\neq0,\pi$ as predicted in Ref. \cite{Buzdin}. The $\varphi$-phase, different from the other phases, is a doubly degenerate state which possesses two energy minima at $\phi=\pm\varphi$ \cite{Tanaka,Tanaka2,Yip,Ilichev,Goldobin3,Lipman,Goldobin4}. Several schemes to realize the $\varphi_{0}$ phase and the $\varphi$ phase have been proposed \cite{Alidoust, Goldobin, Goldobin2}. For example, it is expected that the later phase can be realized in periodic alternating $0$ and $\pi$ junction structures \cite{Buzdin2}; recently the evidence of the phase has been found experimentally\cite{Sickinger}. Actually,  Josephson junctions can also host the mixture of the states, such as $\varphi_{0}\pm\varphi$ phases proposed by E. Goldobin et. al \cite{Goldobin} more recently.

The formation of rich phases in Josephson junctions is based on the current-phase relations (CPRs). Generally, the Josephson current can be expressed as the composition of the harmonics $\sin{n\phi}$ and $\cos{n\phi}$, in which the integer number $n$ denotes the $n$th order contribution. It is demonstrated that the lowest order current (LOC) with $n=1$, $\sin{\phi}$ or $\cos{\phi}$, is absent in spin-singlet superconductor$\vert$spin-triplet superconductor junctions due to the orthogonality of the Cooper pair wave functions \cite{Pals}. However, the situation will be changed as predicted in Ref.~\cite{Pals} when the interface is magnetically active. For example, the interface which is a ferromagnetic barrier or of spin-orbit coupling can lead to the Josephson current proportional to $\cos{\phi}$ when the triplet superconductor is in the chiral $p$-wave state \cite{Tanaka4,Asano}. Furthermore, the dependence of CPRs on the magnetization in the barrier can bring different phases in spin-triplet Josephson junctions with $p$-wave paring. The $0$-$\pi$ transition has been found when superconductor is characterized by the {\bf{d}}-vector with a uniform direction \cite{Brydon01,Brydon02}. Nevertheless, since the direction of the ${\bf{d}}$-vector is independent of wavevectors, more phases cannot be expected in the junctions although the interplay between ferromagnetism and triplet superconductivity can give many interesting and important physical results \cite{Bujnowski,Brydon03,Brydon04}.

In this paper, we propose a concise scheme to realize rich ground states in Josephson junctions consisting of helical $p$-wave superconductors (HPSs) with paring symmetries $k_{x}\hat{x}\pm k_{y}\hat{y}$ and $k_{y}\hat{x}\pm k_{x}\hat{y}$ and a ferromagnet (F). We are interested in these helical superconducting states for many reasons. The states, with ${\bf{d}}$-vectors pinned in the crystallographic $ab$-plane, are candidates for the paring in Sr$_{2}$RuO$_{4}$ \cite{Mackenzie, Maeno, Zhang} and the triplet part of the order parameter in the noncentrosymmetric superconductor CePt$_{3}$Si \cite{Maeno, Bauer}. Further, $k_{x}\hat{x}+k_{y}\hat{y}$ is the two-dimensional analog of the BW state (B phase) in $^3$He \cite{Mackenzie, Maeno}; $k_{y}\hat{x}-k_{x}\hat{y}$ is analogous to the quantum spin Hall system \cite{Qi}. Recently, new symmetries of charge conductance in F$\vert$HPS junctions \cite{Cheng} and peculiar features of spin accumulation in spin-singlet superconductor$\vert$HPS junctions \cite{Lu} are found. The selection rules for LOC in spin-singlet superconductor$\vert$F$\vert$HPS junctions are also summarized \cite{Cheng2} which are distinct from those in the junctions involving triplet superconductor described by a uniform ${\bf{d}}$-vector \cite{Brydon}. As a result, it is reasonable to expect anomalous Josephson effects in the helical $p$-wave Josephson junctions. How CPRs depend on the orientation of the magnetization and which phases the junctions can host are questions to be answered.

In the present work, we systematically study CPRs and ground states of HPS$\vert$F$\vert$HPS junctions using the method of quasiclassical Green's functions with the generalized Riccati parametrization \cite{Eschrig, Eschrig2}. In order to conveniently describe the anisotropic superconductor in the junctions, we show explicitly the diagrammatic representation of the boundary conditions for the method. Through numerical calculations, we find the junctions can host the $0$ phase, $0+\pi$ phase, $\pi$ phase, $\varphi_{0}$ phase and $\varphi$ phase, where the $0+\pi$ phase is a new ground state in which the free energy has two minima at $\phi=0$ and $\phi=\pi$. The transition from one phase to another can be realized through controlling the direction of the magnetization with a weak external filed. The phase diagrams are presented in the orientation space of the magnetization or in the space spanned by the magnitude of the magnetization and the non-magnetic potential. The selection rules for LOC are derived from CPRs which are responsible for the formation of rich phases. In order to explain the rules, we construct a Ginzburg-Landau type of free energy of the junctions with ${\bf{d}}$-vectors in HPS and the magnetization in F, which reveals the interaction mechanism between the helical $p$-wave superconductivity and ferromagnetism. We also clarify the Andreev bound states (ABS) formed at the interface and the novel symmetries in CPRs.

The paper is organized as follows. In Sec. $2$, we establish the theoretical framework which will be used to obtain the results. In Sec. ${3}$, we present the detailed numerical results for the junctions. The features of CPRs and phase diagrams can be found there. In Sec. ${4}$, we further discuss the selection rules for LOC from the viewpoint of free energy. Sec.
${5}$ concludes the work.
\section{Quasiclassical Green's function formalism}
\begin{figure}[h]
\begin{center}
\includegraphics[width=8cm]{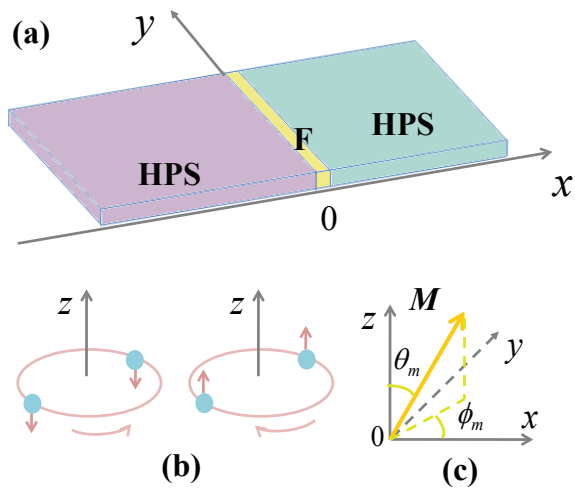}
\end{center}
\caption{(Color online) (a): Schematic illustration of HPS/F/HPS junctions. The $x(y)$-axis is defined by the crystallographic $a(b)$-axis. (b): The spins of Cooper pairs for the helical states in HPSs which can be thought of as the superposition of the two states with spin parallel and anti-parallel to the $z(c)$-axis. The $\bf{d}$-vectors are pinned in the $xy$-plane. (c): The magnetization in F specified by the polar angle $\theta_{m}$ and the azimuthal angle $\phi_{m}$.}
\label{f1}
\end{figure}

We consider the Josephson junctions in the clean limit as shown in fig.~\ref{f1}. The barrier, located at $x=0$ with its interface along the $y$-axis, is modeled by a delta function $U(x)=(U_{0}+{\bf{M}}\cdot\hat{\sigma})\delta(x)$ in which $U_{0}$ and ${\bf{M}}\cdot\hat{\sigma}$ denote the non-magnetic potential and the ferromagnetic term, respectively.
The magnetization ${\bf{M}}=M(\sin{\theta_{m}}\cos{\phi_{m}}, \sin{\theta_{m}}\sin{\phi_{m}}, \cos{\theta_{m}})$ with $\theta_{m}$ the polar angle and $\phi_{m}$ the azimuthal angle which span the orientation space of the magnetization.
For the superconductors, we consider the following helical states,
\begin{equation}
\eqalign{
{\bf{d}}_{1}=\Delta_{0}(k_{x}\hat{x}+k_{y}\hat{y}),\\
{\bf{d}}_{2}=\Delta_{0}(k_{x}\hat{x}-k_{y}\hat{y}),\\
{\bf{d}}_{3}=\Delta_{0}(k_{y}\hat{x}+k_{x}\hat{y}),\\
{\bf{d}}_{4}=\Delta_{0}(k_{y}\hat{x}-k_{x}\hat{y}),}
\label{eq1}
\end{equation}
with $\Delta_{0}$ the temperature-dependent gap magnitude which is determined by the BCS-type equation.
For simplicity, we use HP$_{\alpha}$S to denote the helical $p$-wave superconductor with the ${\bf{d}_{\alpha}}$-vector.

The HPS can be described by the quasiclassical Green's function $\check{g}$, a $2\times2$ matrix in Keldysh space, which is solution of the Eilenberger equation with the normalization condition $\check{g}\otimes \check{g}=-\pi^2\check{1}$. For the physical quantities involved in this paper, it is sufficient to obtain the retard Green's function $\hat{g}^{R}$ which is the upper-left element of $\check{g}$. The retard Green's function $\hat{g}^{R}$, a $4\times4$ matrix in spin$\otimes$particle-hole space, can be written as \cite{Eschrig}
\begin{equation}
\hat{g}^{R}=-2\pi i\left(
\begin{array}{cc}
g&f\\
-\tilde{f}&-\tilde{g}\\
\end{array}
\right)+i\pi\hat{\tau}_{3},
\label{eq2}
\end{equation}
with the parametrization
\begin{equation}
\eqalign{
g=(1-\gamma\tilde{\gamma})^{-1},~~~f=(1-\gamma\tilde{\gamma})^{-1}\gamma,\\
\tilde{g}=(1-\tilde{\gamma}\gamma)^{-1},~~~\tilde{f}=(1-\tilde{\gamma}\gamma)^{-1}
\tilde{\gamma},}
\label{eq3}
\end{equation}
in which $\gamma$ and $\tilde{\gamma}$ are the retard coherence functions. Physically, $\gamma$ ($\tilde{\gamma}$) describes the probability amplitude for conversion of a hole (particle) to a particle (hole). The coherence functions, $2\times2$ matrices in spin space, are a generalization of the so-called Riccati amplitudes.
For simplicity, we have omitted the superscript``R" for the retard functions $g, f, \tilde{g}, \tilde{f}, \gamma$ and $\tilde{\gamma}$.

The coherence functions obey the Riccati-type transport equations
\begin{equation}
\eqalign{
(i\hbar{\bf{v}}_{f}\cdot\nabla+2\varepsilon)\gamma=\gamma\tilde{\Delta}\gamma-\Delta,\\
(i\hbar{\bf{v}}_{f}\cdot\nabla-2\varepsilon)\tilde{\gamma}=\tilde{\gamma}\Delta
\tilde{\gamma}-\tilde{\Delta},}
\label{eq4}
\end{equation}
with the boundary (initial) conditions, which are numerically stable. Here, ${\bf{v}}_{f}$ is the Fermi velocity, $\varepsilon$ the quasiparticle energy measured from the Fermi energy, and $\Delta$ the energy gap matrix with the relation $\tilde{\Delta}({\bf{k}})=[\Delta(-{\bf{k}})]^{*}$. As in Ref.~\cite{Eschrig}, we use in the following $\gamma,\tilde{\gamma}$ and $\Gamma, \tilde{\Gamma}$ to denote the incoming and outgoing quantities, respectively. The quasiclassical Green's function characterized by the Fermi momentum ${\bf{p}}_{f}$ is composed of both incoming and outgoing quantities. The solutions for $\gamma_{1}$, $\tilde{\gamma}_{1}$ and $\gamma_{2}$, $\tilde{\gamma}_{2}$ in the left (subscript $1$) and the right (subscript $2$) superconductor are stable when integrating the equations from the bulk to the interface; the initial conditions are their bulk values in the superconductor (see Appendix A). The solutions for $\Gamma_{1}$, $\tilde{\Gamma}_{1}$ and $\Gamma_{2}$, $\tilde{\Gamma}_{2}$ are stable when integrating the equations from the interface to the bulk; the initial conditions are their values at the interface which can be expressed by the incoming quantities and the scattering matrix $\check{S}$ in the normal state. For example, $\tilde{\Gamma}_{1}$ can be written as
\begin{eqnarray}
\tilde{\Gamma}_{1}=\tilde{\gamma}_{11}+\tilde{\gamma}_{12}(1-\gamma_{2}\tilde{\gamma}_{22})^{-1}\gamma_{2}\tilde{\gamma}_{21},
\label{eq5}
\end{eqnarray}
where the scattering processes are contained in $\tilde{\gamma}_{\alpha\beta}$ with $\alpha,\beta=1,2$.

The scattering matrix $\check{S}$ is diagonal in the particle-hole space, i.e., $\check{S}=\mbox{diag}(\hat{S},\tilde{S})$ with
\begin{equation}
\hat{S}=\left(\begin{array}{cc}
S_{11}&S_{12}\\
S_{21}&S_{22}
\end{array}\right),~~~
\hat{\tilde{S}}=\left(\begin{array}{cc}
\tilde{S}_{11}&\tilde{S}_{12}\\
\tilde{S}_{21}&\tilde{S}_{22}
\end{array}\right).
\label{eq6}
\end{equation}
The $2\times2$ matrices $S_{11}(S_{22})$ and $S_{21}(S_{12})$ in spin space represent the electron reflection in the left (right) metal and the electron transition from the left (right) metal to the right (left) one, respectively. The hole reflection and transition are represented by the matrices $\tilde{S}_{11}(\tilde{S}_{22})$ and $\tilde{S}_{21}(\tilde{S}_{12})$. Generally, $\check{S}$ depends on the direction of the incident particles, such as the scattering at the spin-orbit coupling interface. The explicit expression of the matrix $\check{S}$ for the ferromagnetic interface considered in this paper is given in Appendix A. In the expression, we have defined the effective magnetization magnitude $X=\frac{Mm}{\hbar^2k_{F}}$, the effective non-magnetic potential $Z=\frac{Um}{\hbar^2k_{F}}$ and $k_{x}^{'}=\frac{k_{x}}{k_{F}}$ with $k_{F}$ the Fermi wavevector.

\begin{figure}[h]
\begin{center}
\includegraphics[width=6cm]{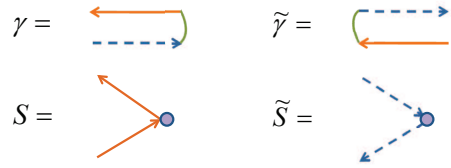}
\end{center}
\caption{(Color online) Diagrammatic symbols of $\gamma,\tilde{\gamma},S$ and $\tilde{S}$. $\gamma$ describes the conversion of a hole (blue dashed line) to a particle (orange solid line); $\tilde{\gamma}$ describes the conversion of a particle to a hole. $S (\tilde{S})$ denotes the scattering of a particle (hole). Note the arrow represents the momentum direction of a particle and the opposite direction of the momentum of a hole.}
\label{f2}
\end{figure}

\begin{figure}[h]
\begin{center}
\includegraphics[width=6cm]{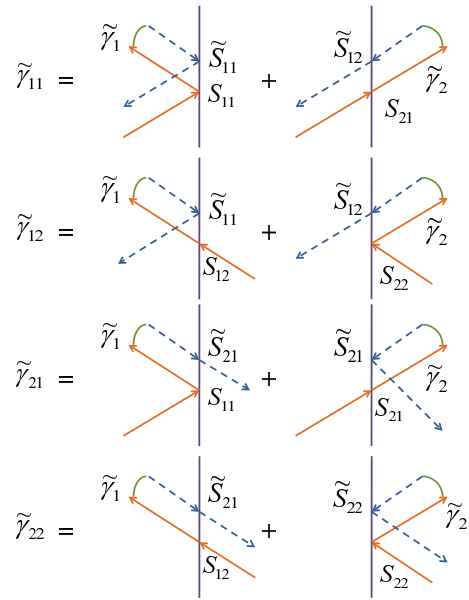}
\end{center}
\caption{(Color online) The scattering processes involved in $\tilde{\gamma}_{\alpha\beta}$ which conserve the momentum component parallel to the interface. $\tilde{\gamma}_{11(22)}$ gives two processes where an incident particle from the left (right) superconductor is converted into a hole moving into the same superconductor. $\tilde{\gamma}_{12(21)}$ gives two processes where an incident particle from the right (left) superconductor is converted into a hole moving into the superconductor on the opposite side.}
\label{f3}
\end{figure}

For anisotropic superconductor, the pair potential and hence the bulk solutions of $\gamma_{1(2)}$ and $\tilde{\gamma}_{1(2)}$ are also dependent on the direction of the momentum of the quasiparticles. In order to show clearly the scattering processes at the interface and to write conveniently and correctly the momentum-dependent quantities, it is necessary to give explicitly the diagrammatic representation of $\gamma_{\alpha\beta}$ and $\tilde{\gamma}_{\alpha\beta}$, in which the directions of the momenta contained in the coherence functions and the scattering matrices are specific. We adopt the diagrammatic symbols for $S, \tilde{S}, \gamma$ and $\tilde{\gamma}$ defined in Ref.~\cite{Eschrig} as shown in fig.~\ref{f2}. The diagrams for $\tilde{\gamma}_{\alpha\beta}$ are given in fig.~\ref{f3}. For simplicity, we do not show the diagrams for $\gamma_{\alpha\beta}$ which can be given in a similar way. Along the reverse direction of the arrow, we can write the expressions of $\tilde{\gamma}_{\alpha,\beta}$,
\begin{equation}
\eqalign{
\tilde{\gamma}_{11}=\tilde{S}_{11}\tilde{\gamma}_{1}S_{11}+\tilde{S}_{12}\tilde{\gamma}_{2}S_{21},\\
\tilde{\gamma}_{12}=\tilde{S}_{11}\tilde{\gamma}_{1}S_{12}+\tilde{S}_{12}\tilde{\gamma}_{2}S_{22},\\
\tilde{\gamma}_{21}=\tilde{S}_{21}\tilde{\gamma}_{1}S_{11}+\tilde{S}_{22}\tilde{\gamma}_{2}S_{21},\\
\tilde{\gamma}_{22}=\tilde{S}_{21}\tilde{\gamma}_{1}S_{12}+\tilde{S}_{22}\tilde{\gamma}_{2}S_{22}.}
\label{eq7}
\end{equation}

Ignoring proximity effect, the retard Green's function $\hat{g}^{R}$ in the left superconductor can be obtained by substituting $\gamma_{1}, \tilde{\Gamma}_{1}$ into eq.~(\ref{eq3}). The Josephson current density can be found from
\begin{equation}
\label{eq8}
J=-eN(0)k_{B}T\sum\limits_{w_{n}}\langle v_{Fx}j(w_{n},\theta)\rangle_{\mbox{FS}_{+}},
\end{equation}
with $j(w_{n},\theta)=\mbox{Tr}[\hat{\tau}_{3}\hat{g}^{R}]$. $N(0)$ is the density of states at the Fermi level in the normal state; the Fermi surface average is only
over positive directions. The Matsubara frequency $w_{n}=2\pi k_{B}T(n+\frac{1}{2})$ with $n$ an integer number and $\theta$ is the angle between the normal to the interface and the momentum of the incident particle. The dimensionless Josephson current denoted by $I_{J}$ can be expressed as $I_{J}=\frac{eIR_{N}}{k_{B}T_{C}}$, where $I=jA$ is the current for junctions with interface area $A$, $R_{N}$ the resistance for junctions in the normal state and $T_{C}$ the critical temperature of superconductor.

\section{Results and discussion}
\subsection{HP$_{1}$S$\vert$F$\vert$HP$_{2}$S junction}
\begin{figure}[h]
\begin{center}
\includegraphics[height=8cm,width=6cm]{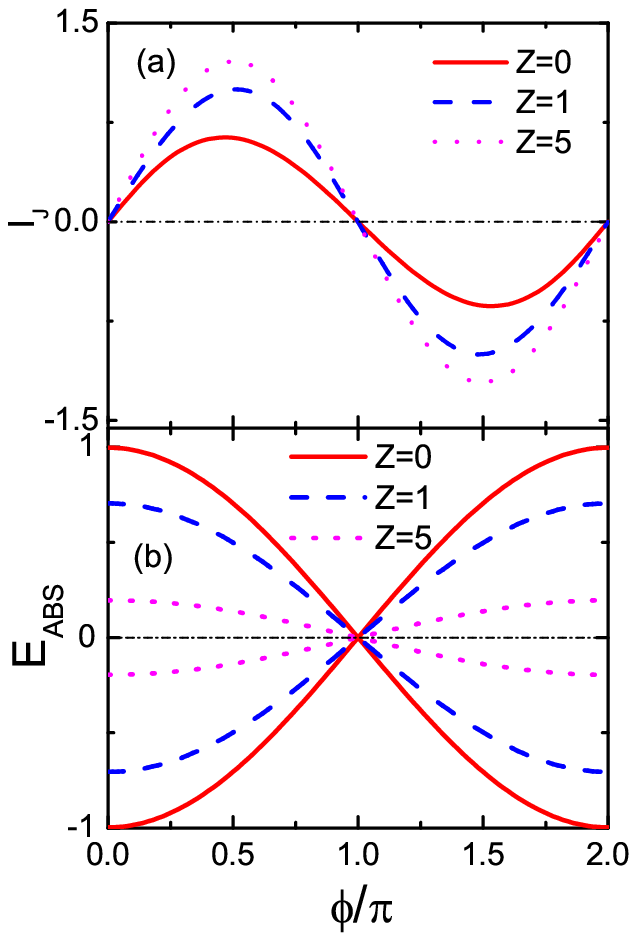}
\end{center}
\caption{(Color online) (a): The CPRs of the HP$_{1}$S$\vert$HP$_{2}$S junction for $X=0$ with $Z=0,1$ and $5$. (b): The corresponding energies of ABS for $\theta=0$.}
\label{f4}
\end{figure}

In our calculations, the temperature is taken as $T=0.3T_{C}$. Firstly, we consider the CPRs for $X=0$. There is no magnetic potential in the interfacial barrier. The HP$_{1}$S$\vert$F$\vert$HP$_{2}$S junction degenerates into the HP$_{1}$S$\vert$HP$_{2}$S one. The effective $j(w_{n},\theta)$ in this case can be written as
\begin{equation}
j(w_{n},\theta)=\frac{8\pi k_{x}^{'2}\gamma^2\sin{\phi}}
{Z^2(1+\gamma^2)^2+k_{x}^{'2}(1+\gamma^4)-2k_{x}^{'2}\gamma^2\cos{\phi}},
\end{equation}
with $\gamma$ defined in Appendix A, which gives the sinusoidal form of the CPRs as shown in fig.~\ref{f4}(a) with $Z=0, 1$ and $5$. When writing the effective expression of $j(w_{n},\theta)$, we have used the relation $\gamma^{*}(w_{n})=1/\gamma(-w_{n})$ with $\gamma^{*}$ the complex conjugate of $\gamma$ and cancelled the terms which have no contribution to the current density $J$. The critical current for the tunneling limit with $Z=5$ is larger than that for the transparent limit with $Z=0$. The dependence of the critical current on the barrier height is different from that of the $s$-wave Josephson junction which also possesses the sinusoidal CPR but the suppressed critical current with increasing $Z$ \cite{Golubov}. For the $s$-wave situation, the energies of ABS are $E=\pm\Delta_{0}\sqrt{1-D\sin^{2}{\phi/2}}$
with $D$ the transmission coefficient, which applies to the point contact or short junction \cite{Bagwell}. The zero-energy level appears when $D=1$ for the transparent limit and will disappear for $D<1$. However, this is not the case for the  HP$_{1}$S$\vert$HP$_{2}$S junction as shown in fig.~\ref{f4}(b). When $\theta=0$, the zero-energy level always exists irrespective of the barrier height. The energies of ABS can be expressed as $E=\pm\Delta_{0}\sqrt{D}\cos{\phi/2}$ with $D=\frac{1}{1+Z^2}$, the transition coefficient for the normal incidence of the quasiparticles, which is just the square of the modulus of the diagonal element of $S_{12}$ or $S_{21}$.

\begin{figure}[h]
\begin{center}
\includegraphics[height=8cm,width=10cm]{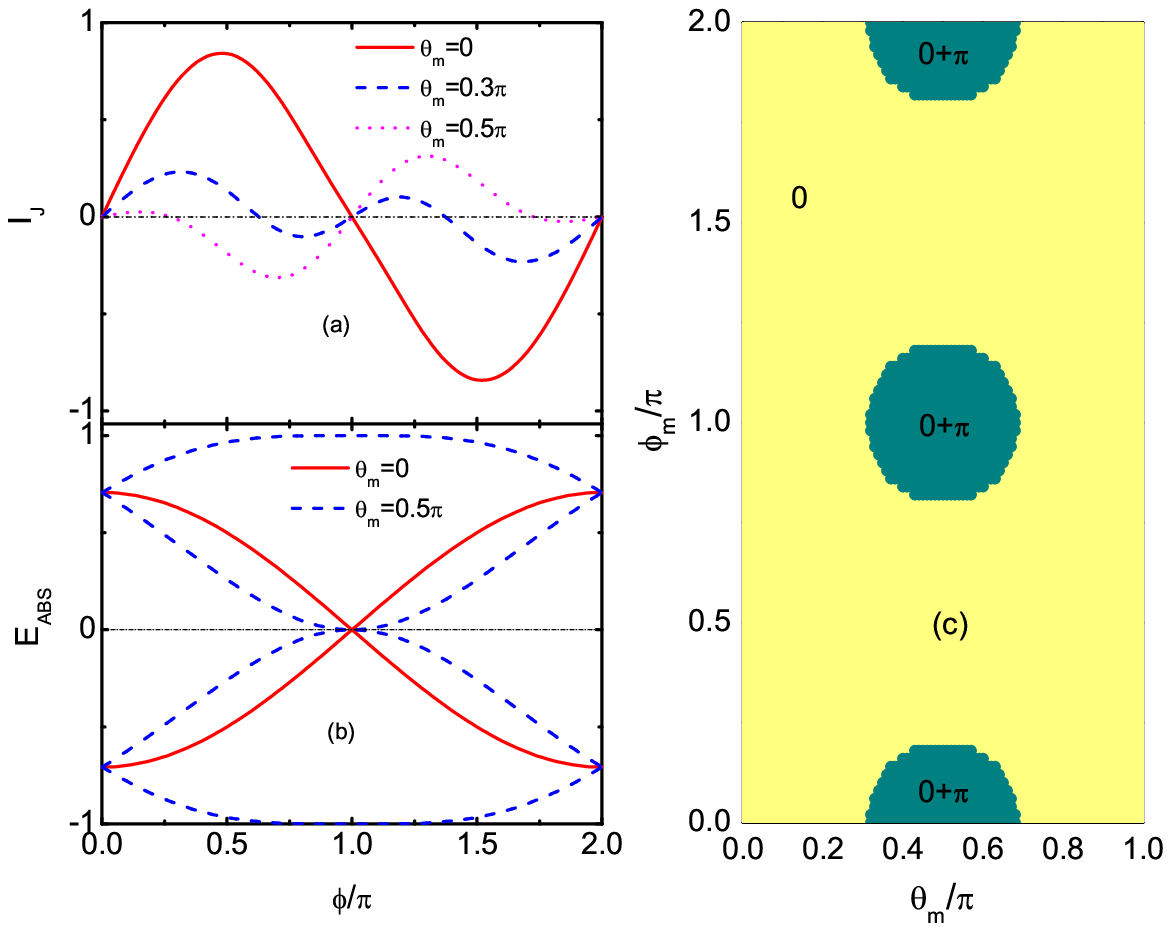}
\end{center}
\caption{(Color online) (a): The CPRs of the HP$_{1}$S$\vert$F$\vert$HP$_{2}$S junction for $Z=0$, $X=1$ and $\phi_{m}=0$. (b): The corresponding energies of ABS. (c): The phase diagram for $0$ phase and $0+\pi$ phase in the orientation space at $Z=0$ and $X=1$.}
\label{f5}
\end{figure}

When $X\ne0$, the CPR strongly depends on the orientation of the magnetization. Fig.~\ref{f5}(a) gives the CPRs for $X=1$ and $Z=0$. We take the azimuthal angle $\phi_{m}=0$. For $\theta_{m}=0$ (${\bf{M}}\parallel\hat{z}$), we have the $\sin{\phi}$-dominated CPR. The free energy of the junction, given by $\frac{\hbar}{2e}\int_{0}^{\phi}{I_{J}(\psi)d\psi}$ , has a minimum at $\phi=0$ with no current across the junction. When the relative angle between the magnetization and the $z$-axis is increased, such as $\theta_{m}=0.3\pi$ or $0.5\pi$ (${\bf{M}}\perp\hat{z}$), the current curve crosses the horizontal line with $I_{J}=0$ at a position in between $\phi=0$ and $\phi=\pi$. The free energy-phase relation has two minima at $\phi=0$ and $\phi=\pi$; the junction is in the $0+\pi$ phase. The energies of ABS for $\theta_{m}=0$ and $\theta_{m}=0.5\pi$ are presented in fig.~\ref{f5}(b). The presence of the $x$-component of the magnetization leads to the splitting of the energies.

From fig.~\ref{f5}(a), we can find the rotation of the magnetization can tune the HP$_{1}$S$\vert$F$\vert$HP$_{2}$S junction between two states: the $0$ phase in which the free energy minimum is obtained at $\phi=0$ and the $0+\pi$ phase in which the free energy minima are obtained at $\phi=0$ and $\pi$. For clarity, we show in fig.~\ref{f5}(c) the phase diagram for the states in the orientation space of the magnetization. There are two characteristics: (a) The $0+\pi$ phase can be realized in two circle-like zones whose centers are located at the points $(\theta_{m},\phi_{m})=(0.5\pi,0)$ and $(0.5\pi,\pi)$, respectively. The ``diameter" of the zones is about $0.36\pi$ long. (b) The phase diagram is symmetric about the axes $\theta_{m}=0.5\pi, \phi_{m}=0.5\pi$ and $\phi_{m}=\pi$ which is a reflection of the symmetries of the CPRs about the direction of the magnetization. They are $I_{J}(\theta_{m},\phi_{m})=I_{J}(\pi-\theta_{m},\phi_{m})=I_{J}(\theta_{m},\pi-\phi_{m})=I_{J}(\theta_{m},\pi+\phi_{m})$. It is interesting to compare the CPRs with those of spin-triplet Josephson junctions characterized by ${\bf{d}}$-vectors with uniform directions \cite{Brydon2}. There, when the ${\bf{d}}$-vectors are both along the $z$-axis, $I_{J}$ is independent of the azimuthal angle of the magnetization. As a result, the orientation space will be divided into rectangular zones by different phases.

\begin{figure}[h]
\begin{center}
\includegraphics[height=8cm,width=10cm]{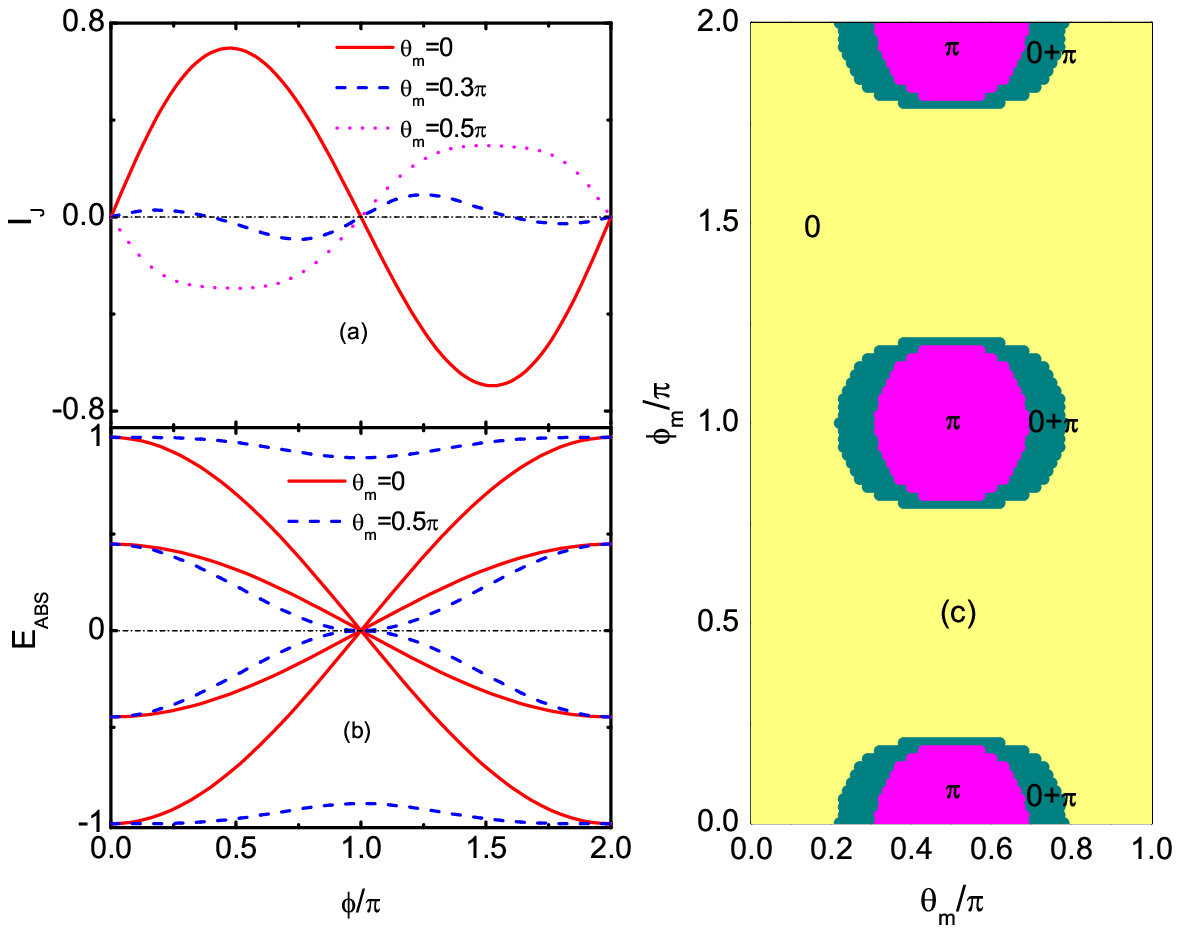}
\end{center}
\caption{(Color online) (a): The CPRs of the HP$_{1}$S$\vert$F$\vert$HP$_{2}$S junction for $Z=1$, $X=1$ and $\phi_{m}=0$. (b): The corresponding energies of ABS. (c): The phase diagram for $0$ phase, $0+\pi$ phase and $\pi$ phase in the orientation space at $Z=1$ and $X=1$.}
\label{f6}
\end{figure}

\begin{figure}[h]
\begin{center}
\includegraphics[height=8cm,width=10cm]{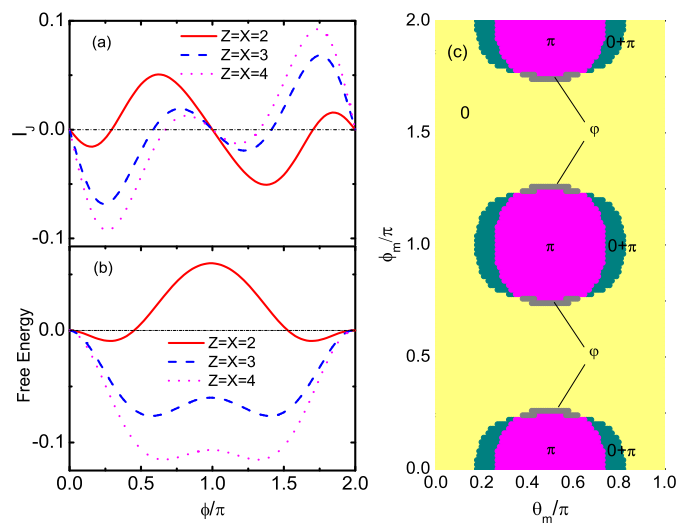}
\end{center}
\caption{(Color online) (a): The CPRs of the HP$_{1}$S$\vert$F$\vert$HP$_{2}$S junction for $Z=X$ when $\phi_{m}=0.25\pi$ and $\theta_{m}=0.5\pi$. (b): The corresponding free energy-phase relations. (c) The phase diagram for $0$ phase, $0+\pi$ phase, $\pi$ phase and $\varphi$ phase in the orientation space at $Z=X=3$.}
\label{f7}
\end{figure}

Now, we turn to the CPRs for $X\neq0$ and $Z\neq0$. Fig.~\ref{f6}(a) shows the currents with $\phi_{m}=0$ at $X=1$ and $Z=1$. The CPR for $\theta_{m}=0.5\pi$, see fig.~\ref{f5}(a), evolves into the $\sin{\phi}$-dominated line shape with a negative critical current, see fig.~\ref{f6}(a), as the non-magnetic potential $Z$ increases from $0$ to $1$. The free energy of the junction in this case has a minimum at $\phi=\pi$; the junction is in the $\pi$ phase. The energy of ABS for $\theta_{m}=0.5\pi$ is given in fig.~\ref{f6}(b). From the phase diagram in fig.~\ref{f6}(c), we can find the zones for the $\pi$ state are located in the ellipse-like zones for the $0+\pi$ state. They possess the same centers: the ``diameter" of the $\pi$ zones is about $0.36\pi$ long; the ¡°major (minor) axis¡± of the $0+\pi$
zones is about $0.54\pi$ $(0.4\pi)$ long. If one continues to increase the values of the $X$ and $Z$ and simultaneously keeps $X=Z$, another new state will emerge at the upper and the lower edges of the $0+\pi$ zones. Fig.~\ref{f7}(a) and (b) plot the CPRs and the free energies for the edge point $(\theta_{m},\phi_{m})=(0.5\pi,0.25\pi)$ at various values of $X$ and $Z$. As shown in the figures, the energy minima of the new state are realized at the location in between $\phi=0$ and $\phi=\pi$ and its symmetric location in between $\phi=\pi$ and $\phi=2\pi$. This new state is the so-called $\varphi$ phase. From fig.~\ref{f7}(b), we can find the locations will tend to $\phi=\pi$ when $X$ and $Z$ are increased. Fig.~\ref{f7}(c) shows the phase diagram for the $0+\pi$, $\pi$ and $\varphi$ phases in the orientation space when $Z=X=3$.

\begin{figure}[h]
\begin{center}
\includegraphics[height=6cm,width=16cm]{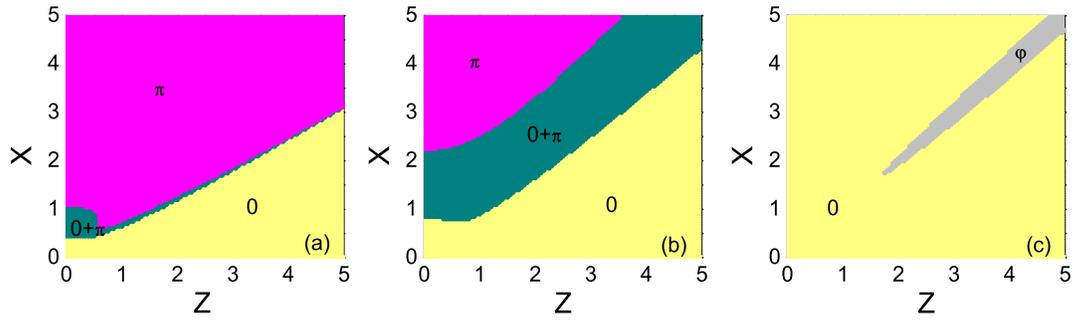}
\end{center}
\caption{(Color online) Phase diagrams for the HP$_{1}$S$\vert$F$\vert$HP$_{2}$S junction in the $Z$-$X$ plane with (a): $\phi_{m}=0$ and $\theta_{m}=0.5\pi$, (b): $\phi_{m}=0$ and $\theta_{m}=0.25\pi$, (c): $\phi_{m}=0.25\pi$ and $\theta_{m}=0.5\pi$.}
\label{f8}
\end{figure}

The magnitude of the magnetization and the non-magnetic potential are important parameters in the realization of different phases. We give in fig.~\ref{f8} the phase diagrams in the $X$-$Z$ plane for three representative points in the orientation space which are denoted by the coordinate $(\phi_{m},\theta_{m})$. Fig.~\ref{f8}(a) is the diagram for the point $(0,0.5\pi)$ which is the center of the zones for the $0+\pi$ and $\pi$ states. When $X<0.4$, one can only obtain the $0$ phase; When $X>3$, one can only obtain the $\pi$ phase. For the moderate values of $X$ with $1<X<3$, the $0+\pi$ phase exists as nearly a boundary line between the $0$ phase and the $\pi$ phase. The parameters $Z$ and $X$ roughly play opposite roles in the formation of the $0$ phase and the $\pi$ phase. This is qualitatively similar to the spin-triplet Josephson junctions with unitary equal-spin pairing states considered in Ref.~\cite{Brydon01}. Fig.~\ref{f8}(b) is the diagram for the point $(0,0.25\pi)$. It is found when the point deviates from the center $(0,0.5\pi)$, the domain of the $\pi$ phase is decreased comparing to that in fig.~\ref{f8}(a). Fig.~\ref{f8}(c) shows the diagram for $(0.25\pi,0.5\pi)$ which is a edge point of the $0+\pi$ phase in fig.~\ref{f7}(c). From the diagram ,we can find the condition for the formation of the $\varphi$ phase which is that $Z$ and $X$ have large enough values (lager than about $2$) and satisfy $Z\approx X$.

Finally, we briefly discuss the presence of the lowest order current, the harmonics $\sin{\phi}$ and $\cos{\phi}$, in the CPRs. There are two main features: (a) The $\sin{\phi}$-type current always exist both for the non-magnetic interface and the magnetic case; (b) No matter how one changes the magnitude of the potentials and the direction of the magnetization, the $\cos{\phi}$-type current will not be obtained. The two features will be further analyzed in Sec. ${4}$.

\subsection{HP$_{1}$S$\vert$F$\vert$HP$_{3}$S junction}
\begin{figure}[h]
\begin{center}
\includegraphics[height=8cm,width=6cm]{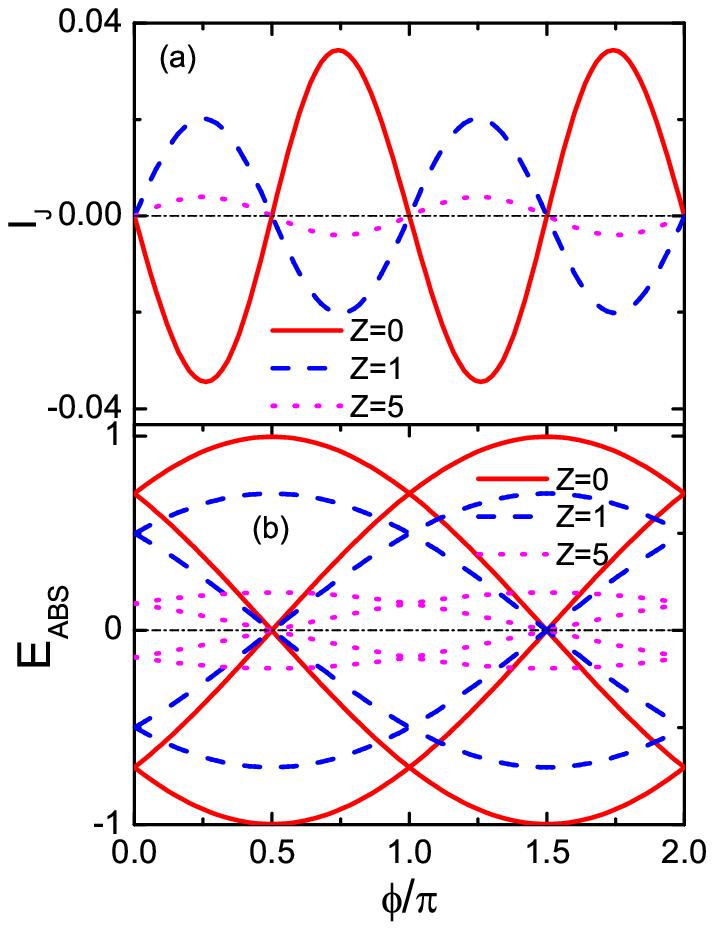}
\end{center}
\caption{(Color online) (a): The CPRs of the HP$_{1}$S$\vert$HP$_{3}$S junction for $X=0$ with $Z=0,1$ and $5$. (b): The corresponding energies of ABS.}
\label{f9}
\end{figure}

For $X=0$, the effective expression of $j(w_{n},\theta)$ can be written as
\begin{eqnarray}
\eqalign{
\fl j(w_{n},\theta)=-4\pi k_{x}^{'2}\cos{\phi}[\frac{\gamma^2}{Z^2(1+\gamma^2)^2+k_{x}^{'2}(1+\gamma^4)-2k_{x}^{'2}\gamma^2\sin{\phi}}\\
-\frac{\gamma^{*2}}{Z^2(1+\gamma^*2)^2+k_{x}^{'2}(1+\gamma^*4)+2k_{x}^{'2}\gamma^*2\sin{\phi}}].}
\end{eqnarray}
The CPRs are shown in fig.~\ref{f9}(a) with $Z=0, 1$ and $5$. Different from the HP$_1$S$\vert$HP$_2$S junction, there is no LOC in the junction HP$_{1}$S$\vert$HP$_{3}$S. The current with the $\sin{2\phi}$ form dominates the CPRs. The energies of ABS with $\theta=0$ are given by $E=\pm\Delta_{0}\sqrt{(1\pm\sin{\phi})/2(1+Z^2)}$ as shown in fig.~\ref{f9}(b). We remind that $I_{J}\propto\sin{2\phi}$ is the typical CPR for spin-singlet superconductor$\vert$spin-triplet superconductor junctions. The absence of LOC in these junctions originates from the orthogonality of the order parameters. For the junctions with the chiral $p$-wave state in triplet superconductor \cite{Tanaka4}, the energies of ABS are given by $E=\pm\Delta_{0}\sqrt{(1+Z^2\pm\sqrt{(1+Z^2)^2-\sin{\phi}^2})/2(1+Z^2)}$.

\begin{figure}[h]
\begin{center}
\includegraphics[height=8cm,width=10cm]{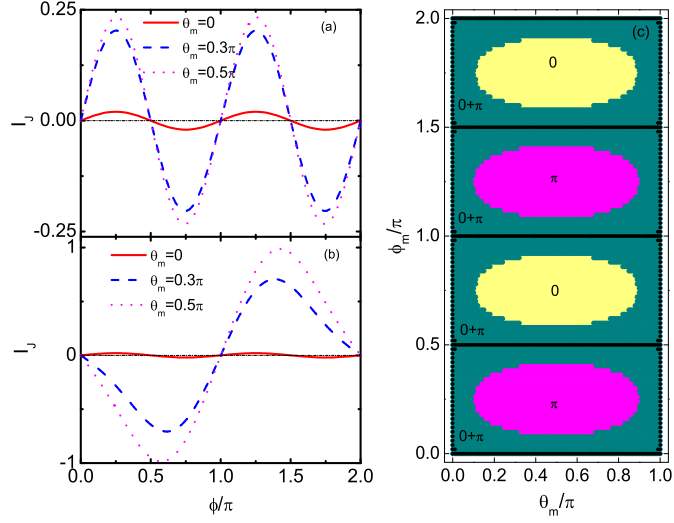}
\end{center}
\caption{(Color online) (a): The CPRs of the HP$_{1}$S$\vert$F$\vert$HP$_{3}$S junction for $Z=0$, $X=1$ and $\phi_{m}=0$. (b): The CPRs for $Z=0$, $X=1$ and $\phi_{m}=0.25\pi$. (c): The phase diagram for $0$ phase, $0+\pi$ phase and $\pi$ phase in the orientation space at $Z=0$ and $X=1$.}
\label{f10}
\end{figure}

Fig.~\ref{f10}(a) plots the CPRs for $X=1$ and $Z=0$. For $\phi_{m}=0$, the variation of the polar angle $\theta_{m}$ only changes the value of the critical current; the CPRs keep the $\sin{2\phi}$ form. That is to say, when $\bf{M}$ is in the $xz$-plane, one cannot expect the presence of LOC. The situation will be changed when $\phi_{m}$ deviates from $0$ as given in fig.~\ref{f10}(b) with $\phi_{m}=0.25\pi$. As $\theta_{m}$ is increased from zero, the harmonic $\sin{\phi}$ emerges and soon dominates the CPR. The junction changes its state from the $0+\pi$ phase to the $\pi$ phase accordingly. The phase diagram in the orientation space is presented in fig.~\ref{f10}(c) which is invariant under a reflection about $\theta_{m}=0.5\pi$ or under a $\pi$ translation of $\phi_{m}$. The invariances of the diagram are the results of symmetries of the current, i.e., $I_{J}(\theta_{m},\phi_{m},\phi)=I_{J}(\pi-\theta_{m},\phi_{m},
\phi)$ and $I_{J}(\theta_{m},\phi_{m},\phi)=I_{J}(\theta_{m},\pi+\phi_{m},\phi)$. In addition to the $0$, $0+\pi$ and $\pi$ phases, there are several black lines with $\theta_{m}=n\pi$ or $\phi_{m}=n\pi/2$ ($n$ is an integer number) in the diagram. For these values, the term $\sin{\phi}$ is absent in the current and $I_{J}\propto\sin{2\phi}$ as shown in fig.~\ref{f10}(a).

\begin{figure}[h]
\begin{center}
\includegraphics[height=9cm,width=16cm]{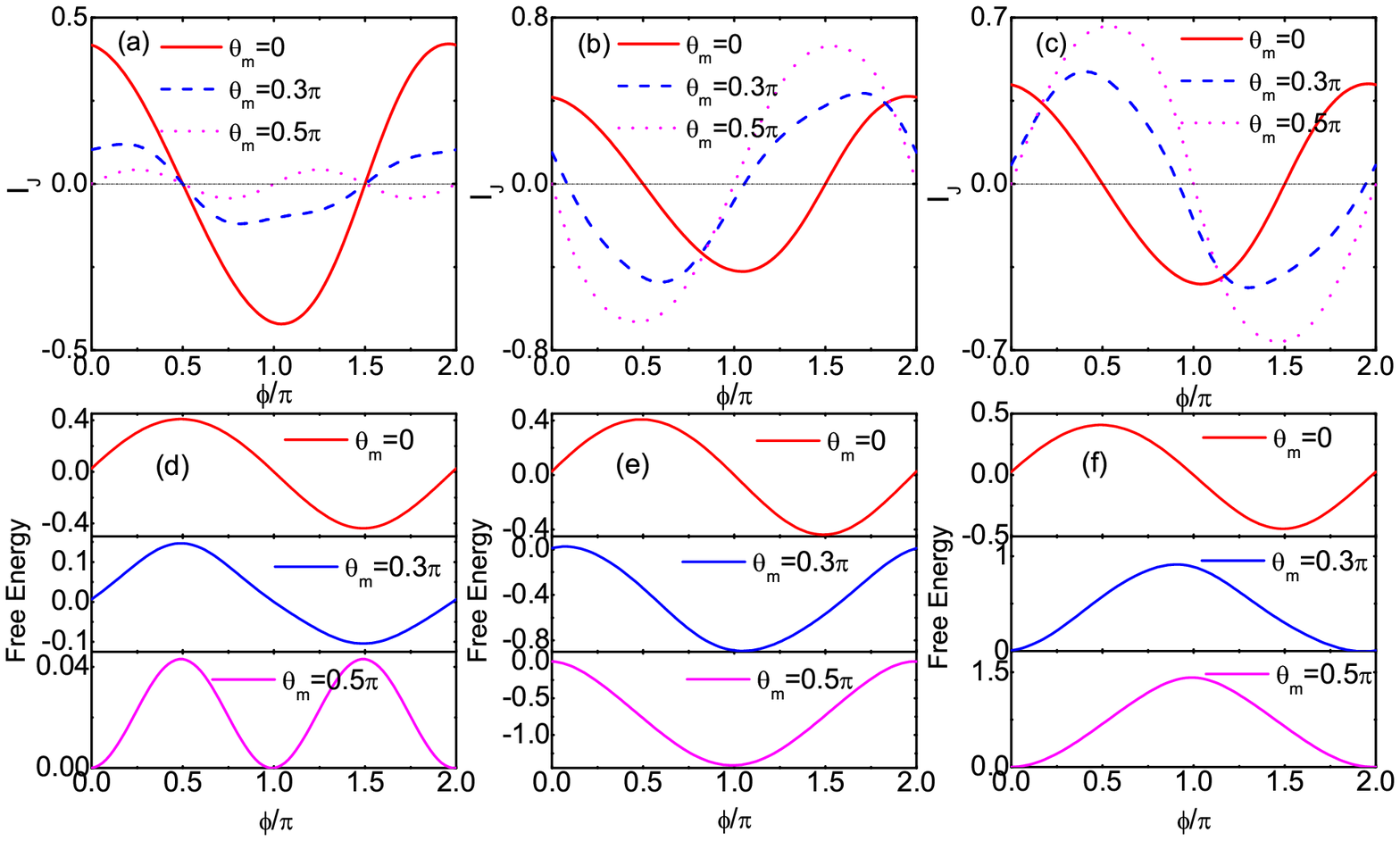}
\end{center}
\caption{(Color online) The CPRs of the HP$_{1}$S$\vert$F$\vert$HP$_{3}$S junction with $Z=1$ and $X=1$ for (a): $\phi_{m}=0$, (b): $\phi_{m}=0.25\pi$ and (c): $\phi_{m}=0.75\pi$. The corresponding free energy are presented in (d)-(f), respectively.}
\label{f11}
\end{figure}

The CPRs for $X\neq0$ and $Z\neq0$ are presented in fig.~\ref{f11} with $X=1$ and $Z=1$. It is found from fig.~\ref{f11}(a) that for $\theta_{m}=0$, LOC with the harmonic $\cos{\phi}$ dominates the CPR. The corresponding free energy has a single minimum at $\phi\approx1.5\pi$ as given in fig.~\ref{f11}(d), which indicates the junction is in the so-called $\varphi_{0}$ phase. As $\theta_{m}$ is increased, LOC is weakened and will disappear when $\theta_{m}=0.5\pi$. The junction changes its state from the $\varphi_{0}$ phase to the $0+\pi$ phase accordingly. For $\phi_{m}=0.25\pi$ in fig.~\ref{f11}(b), as $\theta_{m}$ is increased to $0.5\pi$, $I_{J}\propto\sin{\phi}$ with negative critical current will dominate the CPR. The junction changes its state from the $\varphi_{0}$ phase to the $\pi$ phase accordingly as shown in fig.~\ref{f11}(e). In contrast, for $\phi_{m}=0.75\pi$, $I_{J}\propto\sin{\phi}$ with positive critical current will dominate the CPR when $\theta_{m}$ is increased to $0.5\pi$. The junction changes its state from the $\varphi_{0}$ phase to the $0$ phase accordingly as shown in fig.~\ref{f11}(f). For $Z=1$ and $X=1$, we also have symmetries of $I_{J}$ such as $I_{J}(\theta_{m},\phi_{m},\phi)=-I_{J}(\pi-\theta_{m},\phi_{m},2\pi-\phi)$, $I_{J}(\theta_{m},\phi_{m},\phi)=I_{J}(\theta_{m},\pi+\phi_{m},\phi)$ and $I_{J}(\theta_{m},n\pi/2,\phi)=I_{J}(\theta_{m},(n+1)\pi/2,\phi)$. From the numerical results, we find the $\varphi_{0}$ phase can exist in the HP$_{1}$S$\vert$F$\vert$HP$_{3}$S junction except for $\theta_{m}=0.5\pi$. It is worth noting that the phase can not be achieved in HP$_{1}$S$\vert$F$\vert$HP$_{2}$S junction due to the absence of the $\cos{\phi}$-type current in their CPRs.

From the above results, we can summarize the features of CPRs in the HP$_{1}$S$\vert$F$\vert$HP$_{3}$S junction which are as follows: (a) When $X=0$, LOC is absent. (b) For $X\neq0$, on can obtain the $\sin{\phi}$-type current as long as $\phi_{m}\neq n\pi/2$ and $\theta_{m}\neq n\pi$. (c) For $X\neq0$ and $Z\neq0$, one can obtain the $\cos{\phi}$-type current as long as $\theta_{m}\neq0.5\pi$. We will give the physical explanations of the features in Sec. ${4}$.

\subsection{HP$_{1}$S$\vert$F$\vert$HP$_{4}$S junction}
\begin{figure}[h]
\begin{center}
\includegraphics[height=6cm,width=7cm]{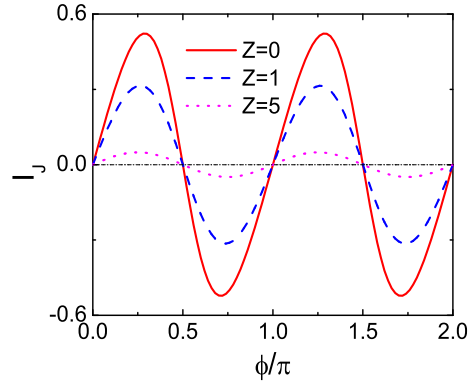}
\end{center}
\caption{(Color online) The CPRs of the HP$_{1}$S$\vert$HP$_{4}$S junction for $X=0$ with $Z=0,1$ and $5$.}
\label{f12}
\end{figure}

The CPRs for $X=0$ are presented in fig.~\ref{f12}, which also satisfy $I_{J}\propto\sin{2\phi}$ as those in the HP$_{1}\vert$HP$_{3}$ junction. One cannot obtain LOC when the magnetic potential is absent in the interface. The effective expression of $j(w_{n},\theta)$ is given by
\begin{equation}
j(w_{n},\theta)=\frac{-8\pi k_{x}^{'4}\vert\gamma\vert^4\sin{2\phi}}
{[k_{x}^{'2}(1+\vert\gamma\vert^4)+Z^2\vert1+\gamma^2\vert^2]^2-4k_{x}^{'2}\vert\gamma\vert^4\sin^2{\phi}}.
\end{equation}
We do not show the energies of ABS because they are the same as those for the HP$_{1}$S$\vert$HP$_{3}$S junction.

\begin{figure}[h]
\begin{center}
\includegraphics[height=8cm,width=10cm]{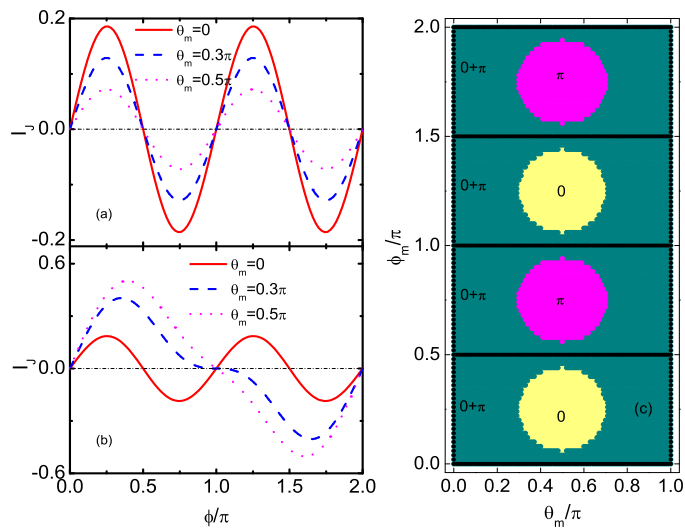}
\end{center}
\caption{(Color online) The CPRs of the HP$_{1}$S$\vert$F$\vert$HP$_{4}$S junction for $Z=0$, $X=2$ and (a): $\phi_{m}$=0; (b): $\phi_{m}=0.25\pi$. The corresponding phase diagram for $0$ phase, $0+\pi$ phase and $\pi$ phase in the orientation space at $Z=0$ and $X=2$.}
\label{f13}
\end{figure}

Fig.~\ref{f13} plots the CPRs for $X=2$ and $Z=0$. For $\phi_{m}=0$, as shown in fig.~\ref{f13}(a), the increment in the value of $\theta_{m}$ only suppresses the critical current. The magnetization in the $xz$-plane will not bring LOC. For $\phi_{m}=0.25\pi$ as shown in fig.~\ref{f13}(b), as $\theta_{m}$ is increased from $0$, the $\sin{\phi}$-type current will soon dominate the CPRs. The junction changes its state from the $0+\pi$ phase to the $0$ phase. Since we have $I_{J}(\theta_{m},\phi_{m},\phi)=-I_{J}(\theta_{m},\pi-\phi_{m},\pi-\phi)$, the harmonic $\sin{\phi}$ with negative critical current will dominate the CPRs for $\phi_{m}=0.75\pi$ when $\theta_{m}$ is increased from $0$. In this case, the junction changes its state from the $0+\pi$ phase to the $\pi$ phase. The phase diagram for $0$ phase, $0+\pi$ phase and $\pi$ phase are presented in fig.~\ref{f13}(c). The symmetries of the diagram are the results of the relations $I_{J}(\theta_{m},\phi_{m},\phi)=I_{J}(\pi-\theta_{m},\phi_{m},\phi)$ and $I_{J}(\theta_{m},\phi_{m},\phi)=I_{J}(\theta_{m},\pi+\phi_{m},\phi)$. There are also some black lines with $\theta_{m}=n\pi/2$ or $\phi_{m}=n\pi$ in the diagram. For these values, we have $I_{J}\propto\sin{2\phi}$ with no LOC.

\begin{figure}[h]
\begin{center}
\includegraphics[height=9cm,width=16cm]{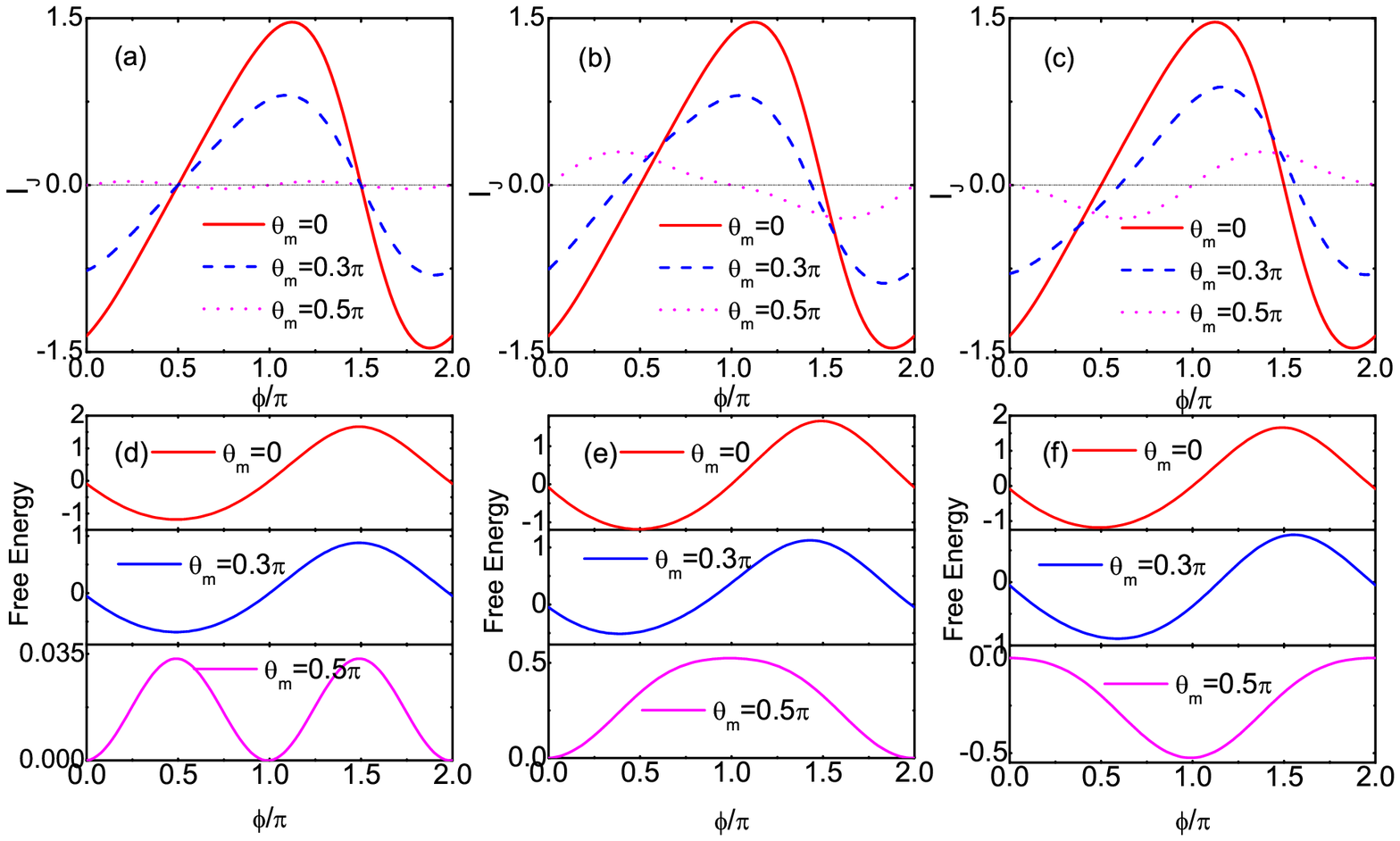}
\end{center}
\caption{(Color online) The CPRs of the HP$_{1}$S$\vert$F$\vert$HP$_{4}$S junction with $Z=2$ and $X=1$ for (a): $\phi_{m}=0$, (b): $\phi_{m}=0.25\pi$ and (c): $\phi_{m}=0.75\pi$. The corresponding free energy are presented in (d)-(f), respectively.}
\label{f14}
\end{figure}

Fig.~\ref{f14}(a)-(c) show the CPRs for $X=1$ and $Z=2$. For $\phi_{m}=0$ in fig.~\ref{f14}(a), the $\cos{\phi}$-type CPR evolves into the $\sin{2\phi}$ form as $\theta_{m}$ is increased. The junction changes its state from the $\varphi_{0}$ phase with $\varphi_{0}\approx0.5\pi$ to the $0+\pi$ phase accordingly as shown in fig.~\ref{f14}(d). However, for $\phi_{m}=0.25\pi$ in fig.~\ref{f14}(b), the $\cos{\phi}$-type CPR will evolve into the $\sin{\phi}$ form as $\theta_{m}$ is increased. The junction changes its state from the $\varphi_{0}$ phase to the $0$ phase accordingly as shown in fig.~\ref{f14}(e). For $\theta_{m}=0.75\pi$ in fig.~\ref{f14}(c), the CPR will evolve into $\sin{\phi}$ form with the negative critical current. The junction changes its state from the $\varphi_{0}$ phase to the $\pi$ phase accordingly as given in fig.~\ref{f14}(f). For $Z=1$ and $X=2$, we have the symmetry relations which are $I_{J}(\theta_{m},\phi_{m},\phi)=I_{J}(\theta_{m},\pi+\phi_{m},\phi)=-I_{J}(\theta_{m},\pi-\phi_{m},\pi-\phi)$ and $I_{J}(\theta_{m},n\pi/2,\phi)=I_{J}(\theta_{m},(n+1)\pi/2,\phi)$. The $\varphi_{0}$ phase can exist in the junction except for $\theta_{m}=0.5\pi$.

The features of CPRs in the HP$_{1}$S$\vert$F$\vert$HP$_{4}$S junction are the same as those in the HP$_{1}$S$\vert$F$\vert$HP$_{3}$S junction which have been summarized in Sec. 3.2. Finally, we discuss briefly the CPRs in other types of helical junctions. For the junctions with the symmetric geometry such as the HP$_{1}$S$\vert$F$\vert$HP$_{1}$S junction, we have trivial CPRs which are dominated by the harmonic $\sin{\phi}$. For other asymmetric junctions, their CPRs can be derived from the junctions we have considered. For example, $I_{J}(\phi)$ in the junction HP$_{2}$S$\vert$F$\vert$HP$_{3}$S is identical to $I_{J}(\pi-\phi)$ in the HP$_{1}$S$\vert$F$\vert$HP$_{4}$S junction.

\section{Free energy and selection rules}
Now, we explain the features of the CPRs of the helical Josephson junctions through constructing the free energy of junctions. The selection rules for LOC will be obtained. Firstly, we consider the non-magnetic junctions with $X=0$. In this case, there are two relevant vectors in each junction, i.e. ${\bf{d}_{1}}$ and ${\bf{d}_{\alpha}}$ with $\alpha=2,3$ or $4$. We calculate the scalar product of the vectors,
\begin{equation}
\eqalign{
\langle{\bf{d}_{1}}\cdot{\bf{d}}_{2}\rangle_{k_{y}}&=\frac{1}{3}\Delta_{0}^{2},\\
\langle{\bf{d}_{1}}\cdot{\bf{d}}_{3}\rangle_{k_{y}}&=0,\\
\langle{\bf{d}_{1}}\cdot{\bf{d}}_{4}\rangle_{k_{y}}&=0,}
\end{equation}
in which $\langle\cdot\cdot\cdot\rangle_{k_{y}}$ denotes the average over the momentum parallel to the interface. The vanishing of the average value implies the ``orthogonality'' of the superconducting states. As a result, LOC will be absent in the non-magnetic junctions HP$_{1}$S$\vert$HP$_{3}$S and HP$_{1}$S$\vert$HP$_{4}$S. In contrast, the harmonic $\sin{\phi}$ dominates the CPR in the HP$_{1}$S$\vert$HP$_{2}$S junction due to the finite average value. This indicates a contribution to the free energy, $\langle{\bf{d}}_{\alpha}\cdot{\bf{d}_{\beta}}\rangle_{k_{y}}\cos{\phi}$, for the non-magnetic junctions. The Josephson current, as the derivative of the free energy with respect to $\phi$, is proportional to $\langle{\bf{d}}_{\alpha}\cdot{\bf{d}_{\beta}}\rangle_{k_{y}}\sin{\phi}$. Hence, the selection rule is just the non-zero condition for the current, i.e. $\langle{\bf{d}}_{\alpha}\cdot{\bf{d}_{\beta}}\rangle_{k_{y}}\neq0$.

Secondly, we consider the magnetic case. There are three relevant vectors, i.e. ${\bf{M}}$, ${\bf{d}_{1}}$ and ${\bf{d}_{\alpha}}$ with $\alpha=2,3$ or $4$, in each junction. In order to include the interaction between the magnetization and the helical superconductivity, we calculate the following scalar product of the vectors,
\begin{equation}
\eqalign{
\langle{\bf{d}}_{1M}\cdot{\bf{d}}_{2M}\rangle_{k_{y}}&=\frac{1}{4}\Delta_{0}^{2}[\frac{1}{3}(7+\cos{\theta_{m}})
-2\sin^2{\theta_{m}}\cos{2\phi_{m}}],\\
\langle{\bf{d}}_{1M}\cdot{\bf{d}}_{3M}\rangle_{k_{y}}&=-\frac{1}{2}\Delta_{0}^{2}\sin^2{\theta_{m}}\sin{2\phi_{m}},\\
\langle{\bf{d}}_{1M}\cdot{\bf{d}}_{4M}\rangle_{k_{y}}&=\frac{1}{6}\Delta_{0}^{2}\sin^2{\theta_{m}}\sin{2\phi_{m}},}
\end{equation}
in which ${\bf{d}}_{\alpha M}$ with $\alpha=1,2,3,4$ denote the $\bf{d}$-vectors written in the spin space of the magnetization which can be obtained by performing unitary transformations (see Appendix B). The averages are determined by the orientation of the magnetization. Since $\langle{\bf{d}}_{1M}\cdot{\bf{d}}_{2M}\rangle_{k_{y}}>0$ holds in the whole space of the orientation, the $\sin{\phi}$-type current always exists in the junction HP$_{1}$S$\vert$F$\vert$HP$_{2}$S irrespective of $\theta_{m}$ and $\phi_{m}$. The condition for $\langle{\bf{d}}_{1M}\cdot{\bf{d}}_{3(4)M}\rangle_{k_{y}}\neq0$ is $\sin{\theta_{m}}\neq0$ and $\sin{2\phi_{m}}\neq0$, therefore one can expect the $\sin{\phi}$-type current when $\theta_{m}\neq n\pi$ and $\phi_{m}\neq n\pi/2$ in the junction HP$_{1}$S$\vert$F$\vert$HP$_{3(4)}$S. This implies a contribution to the free energy, $\langle{\bf{d}}_{\alpha M}\cdot{{\bf{d}}_{\beta M}}\rangle_{k_{y}}\cos{\phi}$, for the magnetic junctions. Accordingly, the Josephson current is proportional to $\langle{\bf{d}}_{\alpha M}\cdot{{\bf{d}}_{\beta M}}\rangle_{k_{y}}\sin{\phi}$.

Thirdly, for the magnetic case, we can also construct another scalar quantity involving both the magnetization $\bf{M}$ and two $\bf{d}$-vectors. The averages of the quantity for different junctions are given by
\begin{equation}
\eqalign{
\langle{\bf{M}}\cdot({\bf{d}}_{1}\times{\bf{d}}_{2})\rangle_{k_{y}}&=0,\\
\langle{\bf{M}}\cdot({\bf{d}}_{1}\times{\bf{d}}_{3})\rangle_{k_{y}}&=\frac{1}{3}\Delta_{0}^{2}\cos{\theta_{m}},\\
\langle{\bf{M}}\cdot({\bf{d}}_{1}\times{\bf{d}}_{4})\rangle_{k_{y}}&=-\Delta_{0}^{2}\cos{\theta_{m}}.}
\end{equation}
For the junction HP$_{1}$S$\vert$F$\vert$HP$_{2}$S, the value of the average is zero for all $\theta_{m}$ and $\phi_{m}$; one cannot find the $\cos{\phi}$-type current in the junction. For the junction HP$_{1}$S$\vert$F$\vert$HP$_{3(4)}$S, the vanishing of the average happens only at $\theta_{m}=\pi/2$; one can obtain the $\cos{\phi}$-type current so long as $\theta_{m}\neq\pi/2$ when $Z\neq0$ and $X\neq0$. This implies another contribution to the free energy, $\langle{\bf{M}}\cdot({\bf{d}}_{\alpha}\times{\bf{d}}_{\beta})\rangle_{k_{y}}\sin{\phi}$, for the magnetic junctions. The term $\langle{\bf{M}}\cdot({\bf{d}}_{\alpha}\times{\bf{d}}_{\beta})\rangle_{k_{y}}\cos{\phi}$ contributes to the Josephson current accordingly. The selection rules for the magnetic case are also the non-zero conditions for the current.

\begin{figure}[h]
\begin{center}
\includegraphics[height=6.5cm,width=6.5cm]{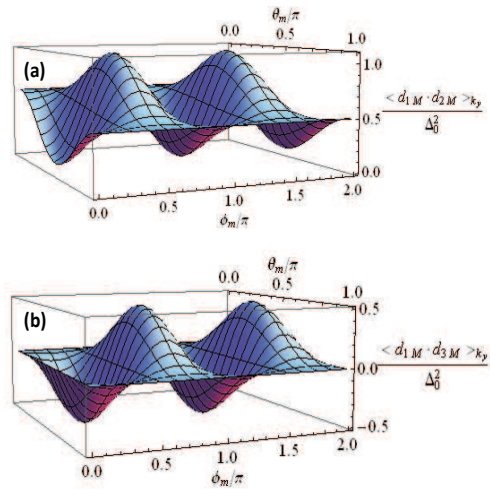}
\end{center}
\caption{(Color online) (a): The normalized $\langle{\bf{d}}_{1M}\cdot{{\bf{d}}_{2M}}\rangle_{k_{y}}$ as a function of $\theta_{m}$ and $\phi_{m}$. (b): The normalized $\langle{\bf{d}}_{1M}\cdot{{\bf{d}}_{3M}}\rangle_{k_{y}}$ as a function of $\theta_{m}$ and $\phi_{m}$.}
\label{f15}
\end{figure}

The complete expressions of the free energy and the Josephson current are very complicated; they are functions of temperature, the non-magnetic potential, the magnitude and the direction of magnetization and the superconducting phase $\phi$. Here, we try to give qualitative explanations of the formation of various phases in helical junctions on the basis of the constructed free energy and the corresponding current. For the HP$_{1}$S$\vert$F$\vert$HP$_{2}$S junction, there is not $\cos{\phi}$-type LOC. The current can be expressed as the composition of $\langle{\bf{d}}_{1M}\cdot{{\bf{d}}_{2M}}\rangle_{k_{y}}\sin{\phi}$ and $\sin{2\phi}$. The second order harmonic $\sin{2\phi}$ originates from the coherent tunneling of even number of Cooper pairs. In this case, the smaller the value of $\langle{\bf{d}}_{1M}\cdot{{\bf{d}}_{2M}}\rangle_{k_{y}}$, the more easily the $0+\pi$ phase comes into being. As shown in fig.~\ref{f15}(a), $\langle{\bf{d}}_{1M}\cdot{{\bf{d}}_{2M}}\rangle_{k_{y}}$ obtains its minimum value at two points in the orientation space of magnetization. The orientation specified by $(\theta_{m},\phi_{m})$ in the zones around the points will lead to the formation of the $0+\pi$ phase which corresponds to the phase diagram given in fig.~\ref{f5}. The $\pi$ phase and the $\varphi$ phase are results of the sign reversal of the current when $X$ or $Z$ is changed. Note, the $\varphi_{0}$ phase does not exist in the junction due to the absence of the $\cos{\phi}$-type LOC.

For the HP$_{1}$S$\vert$F$\vert$HP$_{3}$S junction with $Z=0$, the current is the composition of $\langle{\bf{d}}_{1M}\cdot{{\bf{d}}_{3M}}\rangle_{k_{y}}\sin{\phi}$ and $\sin{2\phi}$. The positive (negative) value of $\langle{\bf{d}}_{1M}\cdot{{\bf{d}}_{3M}}\rangle_{k_{y}}$ is favorable to the formation of the $0$ ($\pi$) phase. $\langle{\bf{d}}_{1M}\cdot{{\bf{d}}_{3M}}\rangle_{k_{y}}$ possess two peaks with the positive maximum value and two valleys with the negative minimum value as shown in fig.~\ref{f15}(b). The values of $\langle{\bf{d}}_{1M}\cdot{{\bf{d}}_{3M}}\rangle_{k_{y}}$ around the peaks and the valleys help to form the $\pi$ phase and the $0$ phase respectively, which leads to the phase diagram in fig.~\ref{f10}. The black lines in the diagram are results of the absence of $\sin{\phi}$ when $\langle{\bf{d}}_{1M}\cdot{{\bf{d}}_{3M}}\rangle_{k_{y}}$=0. For the HP$_{1}$S$\vert$HP$_{3}$S junction with $Z\neq0$, the presence of the $\cos{\phi}$-type LOC for $\theta_{m}\neq\pi/2$ is helpful in the formation of the $\varphi_{0}$ phase. In this situation, the current is the composition of $\langle{\bf{M}}\cdot({\bf{d}}_{1}\times{\bf{d}}_{3})\rangle_{k_{y}}\cos{\phi}$ and $\sin{2\phi}$ when $\theta_{m}=n\pi$ or $\phi_{m}=n\pi/2$ with $\langle{\bf{d}}_{1M}\cdot{{\bf{d}}_{3M}}\rangle_{k_{y}}=0$, or the composition of $\langle{\bf{M}}\cdot({\bf{d}}_{1}\times{\bf{d}}_{3})\rangle_{k_{y}}\cos{\phi}$ and $\sin{\phi}$ when $\theta_{m}\neq n\pi$ and $\phi_{m}\neq n\pi/2$ with $\langle{\bf{d}}_{1M}\cdot{{\bf{d}}_{3M}}\rangle_{k_{y}}\neq0$. The former composition corresponds to the free energy-phase relations in fig.~\ref{f11}(d); the later composition corresponds to the relations in fig.~\ref{f11}(e) and (f). In fig.~\ref{f11}, we have taken $\theta_{m}\leq\pi/2$ which results in the $\varphi_{0}$ phase with $\pi<\varphi_{0}<2\pi$. When $\theta_{m}>\pi/2$, $\langle{\bf{M}}\cdot({\bf{d}}_{1}\times{\bf{d}}_{3})\rangle_{k_{y}}$ will become negative; the value of $\varphi_{0}$ will shift from $\pi<\varphi_{0}<2\pi$ to $0<\varphi_{0}<\pi$. For the HP$_{1}$S$\vert$F$\vert$HP$_{4}$ junction, the explanations are similar to those for the HP$_{1}$S$\vert$F$\vert$HP$_{3}$S junction.

\section{Conclusions}
In this paper, we calculate the current in the helical $p$-wave Josephson junctions using the quasiclassical Green's function technology with the diagrammatic representation of the boundary conditions. Various CPRs are found in the junctions due to the interfacial potential-dependent current which lead to rich phase diagrams. The presence of LOC plays an important role in the formation of different phases. In order to reveal the laws for the occurrence of LOC, we construct two kinds of scalar quantities with the magnetization and the $\bf{d}$-vectors which reflect the interplay of ferromagnetism and helical superconductivity. The non-zero condition for the averages of the quantities will directly lead to the selection rules for LOC. Actually, from our analysis, we can also infer some results for the CPRs in the junctions described by $\bf{d}$-vectors with uniform directions. For example, one will not find LOC in the non-magnetic junctions when two $\bf{d}$-vectors are perpendicular to each other; LOC will not be found in the junctions in which one vector is proportional to $k_{x}$ and the other is proportional to $k_{y}$.

\section*{ACKNOWLEDGMENTS}
This work is supported in part by the National Natural Science Foundation of China (Grant Nos. 11447175, 11547035 and 61572270) and by the Qingdao Science and Technology Program (Grant No. 14-2-4-110-JCH).

\section*{APPENDIX A: Bulk solutions and the scattering matrix}
\setcounter{equation}{0}
\renewcommand{\theequation}{A.\arabic{equation}}
The bulk values of $\gamma_{1}$ and $\tilde{\gamma}_{1}$ in the left superconductor with the HP$_{1}$S-wave symmetry are written as
\begin{equation}
\gamma_{1}=\left(\begin{array}{cc}
\gamma^*&0\\
0&\gamma
\end{array}\right)e^{i\phi},~~~
\tilde{\gamma}_{1}=-\left(\begin{array}{cc}
\gamma^*&0\\
0&\gamma
\end{array}\right)e^{-i\phi},
\end{equation}
with $\gamma=\frac{i\Delta_{0} e^{i\theta}}{w_{n}+\sqrt{w_{n}^2+
\Delta_{0}^2}}$.

The bulk values of $\gamma_{2}$ and $\tilde{\gamma}_{2}$ in the right superconductor are given by
\begin{equation}
\eqalign{
\gamma_{2}&=-\left(\begin{array}{cc}
\gamma^*&0\\
0&\gamma
\end{array}\right),~~~
\tilde{\gamma}_{2}=\left(\begin{array}{cc}
\gamma^*&0\\
0&\gamma
\end{array}\right)~~~\mbox{for HP$_{2}$S},\\
\gamma_{2}&=i\left(\begin{array}{cc}
\gamma^*&0\\
0&-\gamma
\end{array}\right),~~~
\tilde{\gamma}_{2}=i\left(\begin{array}{cc}
\gamma^*&0\\
0&-\gamma
\end{array}\right)~~~\mbox{for HP$_{3}$S},\\
\gamma_{2}&=i\left(\begin{array}{cc}
\gamma&0\\
0&-\gamma^*
\end{array}\right),~~~
\tilde{\gamma}_{2}=i\left(\begin{array}{cc}
\gamma&0\\
0&-\gamma^*
\end{array}\right)~~~\mbox{for HP$_{4}$S}.}
\end{equation}
When we write the expressions, we have taken the directions of wavevectors as shown in fig.~\ref{f3} into account.

For the interface with the ferromagnetic potential, the explicit expressions of the scattering matrices can be given by
\begin{equation}
\label{eqA3}
S_{11}=\left(\begin{array}{cc}
\frac{Z^2-X^2-ik_{x}^{'}(Z-X\cos{\theta_{m}})}{X^2+(k_{x}^{'}+iZ)^2}&
\frac{ik_{x}^{'}X\sin{\theta_{m}}e^{-i\phi_{m}}}{X^2+(k_{x}^{'}+iZ)^2}\\
\frac{ik_{x}^{'}X\sin{\theta_{m}}e^{i\phi_{m}}}{X^2+(k_{x}^{'}+iZ)^2}&
\frac{Z^2-X^2-ik_{x}^{'}(Z+X\cos{\theta_{m}})}{X^2+(k_{x}^{'}+iZ)^2}
\end{array}\right),
\end{equation}
$S_{22}=S_{11}, S_{12}=S_{21}=\hat{1}+S_{11}, \tilde{S}_{11}=\tilde{S}_{22}=S_{11}^*$ and $\tilde{S}_{12}=\tilde{S}_{21}=S_{12}^*$.

\section*{APPENDIX B: The transformation of $\bf{d}$-vectors}
\setcounter{equation}{0}
\renewcommand{\theequation}{C.\arabic{equation}}
The energy gap matrix in the coordinate of spin space in F can be obtained by performing unitary transformation:
\begin{equation}
\Delta_{M}=U^{\dag}\Delta U^{*}
\end{equation}
with
\begin{equation}
U=\left(\begin{array}{cc}
\cos{\frac{\theta_{m}}{2}}e^{-i\phi_{m}/2}&-\sin{\frac{\theta_{m}}{2}}e^{-i\phi_{m}/2}\\
\sin{\frac{\theta_{m}}{2}}e^{i\phi_{m}/2}&\cos{\frac{\theta_{m}}{2}}e^{i\phi_{m}/2}
\end{array}\right).
\end{equation}
Using the relation between the $\bf{d}$-vector and the energy gap matrix given by
\begin{equation}
\Delta=\left(\begin{array}{cc}
-d_{x}+id_{y}&d_{z}\\
d_{z}&d_{x}+id_{y}
\end{array}\right),
\end{equation}
we obtain the vectors ${\bf{d}}_{\alpha M}$ which can be written as
\begin{equation}
\eqalign{
{\bf{d}}_{1M}&=\Delta_{0}[\cos{\theta_{m}}\cos(\theta-\phi_{m})\hat{x}+\sin(\theta-\phi_{m})\hat{y}
+\sin{\theta_{m}}\cos(\theta-\phi_{m})\hat{z}],\\
{\bf{d}}_{2M}&=\Delta_{0}[\cos{\theta_{m}}\cos(\theta+\phi_{m})\hat{x}-\sin(\theta+\phi_{m})\hat{y}
+\sin{\theta_{m}}\cos(\theta+\phi_{m})\hat{z}],\\
{\bf{d}}_{3M}&=\Delta_{0}[\cos{\theta_{m}}\sin(\theta+\phi_{m})\hat{x}+\cos(\theta+\phi_{m})\hat{y}
+\sin{\theta_{m}}\sin(\theta+\phi_{m})\hat{z}],\\
{\bf{d}}_{4M}&=\Delta_{0}[\cos{\theta_{m}}\sin(\theta-\phi_{m})\hat{x}-\cos(\theta-\phi_{m})\hat{y}
+\sin{\theta_{m}}\sin(\theta-\phi_{m})\hat{z}].}
\end{equation}

\section*{Figure Captions}
Figure 1: (Color online) (a): Schematic illustration of HPS/F/HPS junctions. The $x(y)$-axis is defined by the crystallographic $a(b)$-axis. (b): The spins of Cooper pairs for the helical states in HPSs which can be thought of as the superposition of the two states with spin parallel and anti-parallel to the $z(c)$-axis. The $\bf{d}$-vectors are pinned in the $xy$-plane. (c): The magnetization in F specified by the polar angle $\theta_{m}$ and the azimuthal angle $\phi_{m}$.

Figure 2: (Color online) Diagrammatic symbols of $\gamma,\tilde{\gamma},S$ and $\tilde{S}$. $\gamma$ describes the conversion of a hole (blue dashed line) to a particle (orange solid line); $\tilde{\gamma}$ describes the conversion of a particle to a hole. $S (\tilde{S})$ denotes the scattering of a particle (hole). Note the arrow represents the momentum direction of a particle and the opposite direction of the momentum of a hole.

Figure 3: (Color online) The scattering processes involved in $\tilde{\gamma}_{\alpha\beta}$ which conserve the momentum component parallel to the interface. $\tilde{\gamma}_{11(22)}$ gives two processes where an incident particle from the left (right) superconductor is converted into a hole moving into the same superconductor. $\tilde{\gamma}_{12(21)}$ gives two processes where an incident particle from the right (left) superconductor is converted into a hole moving into the superconductor on the opposite side.

Figure 4: (Color online) (a): The CPRs of the HP$_{1}$S$\vert$HP$_{2}$S junction for $X=0$ with $Z=0,1$ and $5$. (b): The corresponding energies of ABS for $\theta=0$.

Figure 5: (Color online) (a): The CPRs of the HP$_{1}$S$\vert$F$\vert$HP$_{2}$S junction for $Z=0$, $X=1$ and $\phi_{m}=0$. (b): The corresponding energies of ABS. (c): The phase diagram for $0$ phase and $0+\pi$ phase in the orientation space at $Z=0$ and $X=1$.

Figure 6: (Color online) (a): The CPRs of the HP$_{1}$S$\vert$F$\vert$HP$_{2}$S junction for $Z=1$, $X=1$ and $\phi_{m}=0$. (b): The corresponding energies of ABS. (c): The phase diagram for $0$ phase, $0+\pi$ phase and $\pi$ phase in the orientation space at $Z=1$ and $X=1$.

Figure 7: (Color online) (a): The CPRs of the HP$_{1}$S$\vert$F$\vert$HP$_{2}$S junction for $Z=X$ when $\phi_{m}=0.25\pi$ and $\theta_{m}=0.5\pi$. (b): The corresponding free energy-phase relations. (c) The phase diagram for $0$ phase, $0+\pi$ phase, $\pi$ phase and $\varphi$ phase in the orientation space at $Z=X=3$.

Figure 8: (Color online) Phase diagrams for the HP$_{1}$S$\vert$F$\vert$HP$_{2}$S junction in the $Z$-$X$ plane with (a): $\phi_{m}=0$ and $\theta_{m}=0.5\pi$, (b): $\phi_{m}=0$ and $\theta_{m}=0.25\pi$, (c): $\phi_{m}=0.25\pi$ and $\theta_{m}=0.5\pi$.

Figure 9: (Color online) (a): The CPRs of the HP$_{1}$S$\vert$HP$_{3}$S junction for $X=0$ with $Z=0,1$ and $5$. (b): The corresponding energies of ABS.

Figure 10: (Color online) (a): The CPRs of the HP$_{1}$S$\vert$F$\vert$HP$_{3}$S junction for $Z=0$, $X=1$ and $\phi_{m}=0$. (b): The CPRs for $Z=0$, $X=1$ and $\phi_{m}=0.25\pi$. (c): The phase diagram for $0$ phase, $0+\pi$ phase and $\pi$ phase in the orientation space at $Z=0$ and $X=1$.

Figure 11: (Color online) The CPRs of the HP$_{1}$S$\vert$F$\vert$HP$_{3}$S junction with $Z=1$ and $X=1$ for (a): $\phi_{m}=0$, (b): $\phi_{m}=0.25\pi$ and (c): $\phi_{m}=0.75\pi$. The corresponding free energy are presented in (d)-(f), respectively.

Figure 12: (Color online) The CPRs of the HP$_{1}$S$\vert$HP$_{4}$S junction for $X=0$ with $Z=0,1$ and $5$.

Figure 13: (Color online) The CPRs of the HP$_{1}$S$\vert$F$\vert$HP$_{4}$S junction for $Z=0$, $X=2$ and (a): $\phi_{m}$=0; (b): $\phi_{m}=0.25\pi$. The corresponding phase diagram for $0$ phase, $0+\pi$ phase and $\pi$ phase in the orientation space at $Z=0$ and $X=2$.

Figure 14: (Color online) The CPRs of the HP$_{1}$S$\vert$F$\vert$HP$_{4}$S junction with $Z=2$ and $X=1$ for (a): $\phi_{m}=0$, (b): $\phi_{m}=0.25\pi$ and (c): $\phi_{m}=0.75\pi$. The corresponding free energy are presented in (d)-(f), respectively.

Figure 15: (Color online) (a): The normalized $\langle{\bf{d}}_{1M}\cdot{{\bf{d}}_{2M}}\rangle_{k_{y}}$ as a function of $\theta_{m}$ and $\phi_{m}$. (b): The normalized $\langle{\bf{d}}_{1M}\cdot{{\bf{d}}_{3M}}\rangle_{k_{y}}$ as a function of $\theta_{m}$ and $\phi_{m}$.

\section*{Figures}

\end{document}